\documentclass[a4paper,12pt]{article}
\usepackage{authblk}
\usepackage{float}
\usepackage{graphicx}
\usepackage{booktabs}
\usepackage{tabularx}
\usepackage{siunitx}
\usepackage{tikz}
\usepackage{amsmath}
\usepackage{amsfonts}
\usepackage{draftwatermark}
\SetWatermarkText{Rough Draft}
\SetWatermarkScale{4}
\SetWatermarkColor{red!25!white}
\usepackage{adjustbox}
\usepackage{multirow}
\usepackage{color}
\usepackage{longtable}
\usepackage{pdflscape}
\usepackage{rotating}
\usepackage{xcolor}
\usepackage{xcolor}
\usepackage[colorlinks = true,
            linkcolor = blue,
            urlcolor  = blue,
            citecolor = blue,
            anchorcolor = blue]{hyperref}
\usepackage[margin=1in,footskip=0.25in]{geometry}
\title{Resonant Structures in $p{}^7\mathrm{Be}$ Scattering and Their Connection to the Astrophysical \textit{S}-Factor\\
}

\author{Anil Khachi$^{1}$\thanks{ anilkhachi1990@gmail.com}
\\\vspace{0.25cm}
$^{1}$Chandigarh Group of Colleges Jhanjeri, Mohali, Punjab, India-140307\\  Chandigarh Engineering College, Department of Applied Science}

\begin{document}
\maketitle
\begin{abstract}
\noindent In this paper, we employ the Variable Phase Approach (VPA) to obtain the scattering phase shifts \( \delta(E, r) \), amplitude function \( A(r) \), and radial wavefunction \( u(r) \) for various channels involved in the astrophysical reaction \( {}^7\mathrm{Be}(p,\gamma)^8\mathrm{B} \). Using the extracted phase shifts, we compute the total and partial cross sections. It is observed that the peaks in the partial cross section correspond to resonant states in the compound nucleus $^8B$, which also manifest as enhancements in the astrophysical \textit{S}-factor. These resonances significantly increase the reaction probability at certain energies, particularly in the low-energy regime relevant to stellar nucleosynthesis. The VPA thus serves as a reliable and efficient method for calculating scattering phase shifts and, in turn, extracting the resonance energies of different partial waves. These resonance energies can provide valuable insight into the energy region where the astrophysical \textit{S}-factor is likely to peak.
\end{abstract}

\noindent \textbf{keywords:} scattering, variable phase approach, phase shift, amplitude function, wave function, S-factor

\section{Introduction}
The astrophysical S-factor is a key quantity in thermonuclear reactions, as it determines the reaction cross-section. It represents the likelihood of specific particle behavior at very low energies. While the \textit{S}-factor can be measured experimentally, this is typically only possible for energies above 100 keV to 1 MeV \cite{1}. However, astrophysical calculations, such as those related to stellar evolution, require values for the S-factor at much lower energies typically around 0.1 to 100 keV, which correspond to core temperatures of stars between $10^6$ K and $10^9$ K.

An example of a radiative capture process is the reaction
\[
p + ^7\text{Be} \rightarrow ^8\text{B} + \gamma
\]
which occurs in the pp-chain and is a primary source of energy for the Sun and most stars in the universe. The unstable nucleus $^8\text{B}$ undergoes weak decay into
\[
^8\text{B} \rightarrow ^8\text{Be} + e^+ + \nu,
\]
eventually leading to two helium nuclei ($^4\text{He} + ^4\text{He}$). The high-energy neutrinos produced from this decay are detectable on Earth and serve as a direct probe of this process happening in the Sun. In the study of various astrophysical phenomena, such as the early universe, stellar evolution, and supernovae, the total cross-section of radiative capture reactions is an important parameter in astrophysical models. Since many of these environments have relatively low temperatures, cross-section measurements need to be made at low energies, ranging from tens to hundreds of keV. While a few of these measurements can be carried out in laboratories, most are obtained by extrapolating to lower energies. Direct measurement of cross-sections for reactions like proton capture involving radioactive nuclei (e.g., $^8\text{B}$ or $^7\text{Be}$) is especially challenging due to their small size and the limited intensity of radioactive beams.

Given these experimental challenges, theoretical models are essential for complementing experimental data. These models can be applied across a range of energies, unaffected by the instability of certain nuclei. However, they too have their difficulties, especially when it comes to analyzing radiative capture reactions involved in stellar nucleosynthesis. A complete nuclear model capable of describing all the reactions relevant to modern stellar evolution is not feasible. Below, we outline the general procedure for calculating the astrophysical S-factor in thermonuclear reactions as also explained by Dubovichenko in detail \cite{1}.

\begin{enumerate}
\item Obtain experimental data for the differential cross-sections or excitation functions $\sigma_{\text{exp}}$ of the elastic scattering of the nuclear particles involved.
\item Perform phase shift analysis of the data, or use previously obtained results, to determine the energy-dependent elastic scattering phase shifts $\delta_\ell(E)$. This is a key step in the calculation of astrophysical S-factors.
\item Construct the interaction potential $V(r)$ that reproduces the extracted phase shifts. This step, known as the potential description of phase shifts, is carried out using the VPA (Variable Phase Approach) \cite{3}.
\item Use the resulting intercluster potentials to calculate the total cross-sections for photodisintegration and radiative capture processes, obtaining the theoretical cross-section $\sigma(E)$ for the photonuclear reaction.
\item Finally, compute the astrophysical S-factor $S(E)$ at low energies for the thermonuclear reaction under study.
\end{enumerate}

All of these steps can be schematically represented in the following form:
\begin{equation*}
\sigma_{exp}\rightarrow \delta_\ell(E)\rightarrow V (r) \rightarrow \sigma(E)\rightarrow S(E)
\end{equation*}
This scheme is identical for all photonuclear reactions and is independent of the reaction energy or some other factors.
In our earlier work we have applied variable phase approach (only potential dependent) for various scattering systems like $n-n$, $n-p$, $n-\alpha$, $n-d$, $\alpha-\alpha$, $\alpha-^3He$  and astrophysically important reaction $\alpha-^3He$ scattering systems where Variational Monte-Carlo methods and genetic algorithms have been employed for the optimization process \cite{4}. We have also obtained $n-p$ quantum mechanical functions like amplitude and wave function for various n-p channels and are in good aggreement to that of Av-18 potential. 

\textit{``This work explores how phase shift information can be used to approximate the astrophysical S-factor, and how the resonance peaks in the total partial cross section correspond to the energy region where the S-factor exhibits significant enhancement''}. The Variable Phase Approach (VPA) is a powerful tool for approximating resonance energies and scattering properties, but it falls short of providing an exact calculation of the S-factor. Instead, the VPA can identify the energy region where the S-factor is expected to peak. This is because the VPA can either compute the interaction potential from the phase shifts or, conversely, calculate the phase shifts given the interaction potential.

If one knows the phase shifts, the partial and total cross-sections can be directly calculated. However, the S-factor depends not only on the cross-section but also on the detailed capture cross-section, which accounts for the reaction of a projectile (like a proton or alpha particle) with a target nucleus to form a compound nucleus ($^8B$). The \textit{S}-factor is related to the astrophysical reaction rate, which is highly sensitive to low-energy resonance behavior. In such reactions, particularly at astrophysical temperatures (typically in the range of 1-100 keV), quantum tunneling and resonance effects dominate.

To compute the S-factor precisely, one needs more than just the phase shift information. Knowledge of the electromagnetic transitions is crucial. This includes the electromagnetic capture cross-section, which accounts for the transitions between nuclear states induced by the electromagnetic interaction (e.g., E1 transitions in radiative capture). The VPA, however, does not provide information about the electromagnetic form factors or the transition rates, which are essential for calculating the capture cross-section.

It is to be noted that the aim of our work is not to optimize the potential parameters to fit the phase shifts. Instead, we use the parameters provided by Dubovichenko \textit{et al.} directly within the Variable Phase Approach (VPA) \cite{1}. Our focus is to investigate whether, by knowing the resonance energies alone (through VPA), one can extract meaningful information relevant to the S-factor.

\section{Methodology:}
\subsection{Potential description of elastic scattering phase shifts}
The phase shifts are extracted from the experimental differential cross sections using phase shift analysis. For the bound states of light nuclei in cluster channels, the potentials are constructed not only on the basis of the description of the scattering phase shifts, but also using certain additional requirements. For example, one such requirement is the reproduction of the binding energy and other basic characteristics of the bound states of the nuclei. We will be using intercluster potentials of simple Gaussian form with a point-like Coulomb term \cite{1}
\begin{equation}
V(r)=-V_0 \text{exp}\{-\alpha r^2\}+1.439975 \frac{Z_1Z_2}{r^2}; ~~~~~~(MeV)
\label{poteq}
\end{equation}
Here $V_0$ is the depth of the potential (in $MeV$) and $\alpha$ is the inverse square of the range (in $fm^{-2}$). In our calculations we have taken the masses $M_{Be-7}$ = 6534.19; $MeV/c^2$ and $M_p$= 938.272046 $MeV/c^2$ respectively. The value of $\hbar$= 197.3269718 $MeV.fm$ was utilized in the calculations. The potential can be used directly inside the VPA equation as discussed in next subsection.
\subsection{Variable Phase Approach (VPA)}  
\noindent The Schr$\ddot{\text{o}}$dinger wave equation for a spinless particle with energy E and orbital angular momentum $\ell$ undergoing scattering is given by
\begin{equation}
\frac{\hbar^2}{2\mu} \bigg[\frac{d^2}{dr^2}+\big(k^2-\ell(\ell+1)/r^2\big)\bigg]u_{\ell}(k,r)=V(r)u_{\ell}(k,r)
\label{Scheq}
\end{equation}
Where $k_{c.m}=\sqrt{E_{c.m}/(\hbar^2/2\mu)}$,  $\hbar^2/2\mu$ = 23.729363 MeV fm$^{2}$, $\ell$ indicates the orbital quantum number, $V(r)$ is the interaction potential (as given in equation \ref{poteq}) and the energy of projectile $E_{lab}$ is converted into center of mass energy $E_{c.m.}$ for non-relativistic kinematics as $E_{c.m}=\frac{M_{Be^7}}{M_{Be^7}+M_p}E_{lab}$. Second order differential equation  Eq.\ref{Scheq} has been transformed to the first order non-homogeneous differential equation of Riccati type given by following equation \cite{3}  
\begin{equation}
\delta_{\ell}'(k,r)=-\frac{V(r)}{k\left(\hbar^{2} / 2 \mu\right)}\bigg[\cos(\delta_\ell(k,r))\hat{j}_{\ell}(kr)-\sin(\delta_\ell(k,r))\hat{\eta}_{\ell}(kr)\bigg]^2
\label{PFMeqn}
\end{equation}
Here Prime denotes differentiation of phase function with respect to distance. Equation \ref{PFMeqn} will be used as our main tool for studying the properties of scattering phase shifts for $p{}^7\mathrm{Be}$ system. The advantage of the variable phase approach is that using the well known traditional approaches, one integrates the radial Schr$\ddot{\text{o}}$dinger equation from the origin to the asymptotic region where the potential $V(r)$ is negligible, and then compares the phase of the radial wave function with that of a free wave and thus obtain the phase shift. In the variable phase approach we need only integrate a first-order nonlinear differential equation from the origin to the asymptotic region, thereby obtaining directly the value of the scattering phase shift. In integral form, the VPA equation or ``phase equation" can be written as
\begin{equation}
\delta(k,r)=-\frac{1}{{k\left(\hbar^{2} / 2 \mu\right)}}\int_{0}^{r}{V(r)}\bigg[\cos(\delta_{\ell}(k,r))\hat{j_{\ell}}(kr)-\sin(\delta_{\ell}(k,r))\hat{\eta_{\ell}}(kr)\bigg]^2 dr
\end{equation}
Eq.\ref{PFMeqn} is numerically solved using Runge-Kutta 5$^{th}$ order (RK-5) method subjected to condition $\delta_{\ell}(0) = 0$. The above equation can be used to evaluate the phase function $\delta(r)$, which defines the total phase shift up to r. Then the phase shift (solution of a non-linear differential equation) is the limiting value
\begin{equation}
    \delta=\lim_{r \to \infty} \delta_\ell(r)
\end{equation}
To calculate $\delta$ the phase function $\delta(k,r)$ is evaluated up to a cut-off point at which $\delta(k,r)$ saturates or stabilizes. The function $\delta_{\ell}(r)$ does have a limit as ${r \to \infty}$ and that limit is the desired extracted phase shift. The physical implication of the cut-off point is that it is the radial distance beyond which the inter-nuclear potential can be considered zero. For $\ell = 0$, the Riccati-Bessel and Riccati-Neumann functions $\hat{j}_0$ and $\hat{\eta}_0$ get simplified as $\sin(kr)$ and $-\cos(kr)$, so Eq.\ref{PFMeqn}, for $\ell = 0$ becomes 
\begin{equation}
\delta'_0(k,r)=-\frac{V(r)}{{k\left(\hbar^{2} / 2 \mu\right)}}\sin^2[kr+\delta_0(k,r)]
\label{zero}
\end{equation}


\begin{table}[ht]
\centering
\caption{The elastic scattering phase shift parameters, as provided by Dubovichenko \textit{et al.}, are presented. The calculated resonance energies for various states are listed in the fifth column and compared with the corresponding values from \cite{1}, which are shown in bold.}
\label{tab:resonances}
\begin{tabular}{@{} c l S[table-format=4.3] S[table-format=1.2] S[table-format=1.2] S[table-format=4.0]@{}}
\toprule
No. & {$^{(2S+1)}L_J$} & {$V_0\ (\mathrm{MeV})$} & {$\alpha\ (\mathrm{fm}^{-2})$} & {Res. energy $E_{\rm cm}\ (\mathrm{MeV})$} \\
\midrule
1  & $^3P_1 (R1)$ & 146.0658 & 0.47 & 0.565~~~~\{\textbf{0.632}\} \\
2  & $^3P_1 (R2)$ & 201.05 & 0.7 &3.586~~~~ \{\textbf{3.16}\}  \\
3 & $^3P_2(Bound~state)$ & 676.160 & 2.30 & 3.376~~~~\{\textbf{2.41}\}  \\
4 & $^3D_2$ & 29.95 & 0.062 & 3.742~~~~\{\textbf{3.66}\} \\
5 & $^3F_3$ & 39.364 & 0.04 & 2.523~~~~\{\textbf{2.18}\} \\
6 & $^3S_1$ & 300.000 & 1.00 & {--} \\
\bottomrule
\end{tabular}
\end{table}

In the above equation, the function $\delta_0(k,r)$ was termed ``Phase function" by Morse and Allis. Because the nucleon-nucleon interactions are short range, we integrate our phase shift equations to a distance of about 5 fermis where the nucleon-nucleon potential becomes very weak. The equation for  amplitude function with the initial condition $A_{\ell}(0)=1$ is obtained in the form

\begin{equation}
\begin{aligned}
    A_{\ell}^{\prime}(r) = &-\frac{A_{\ell} V(r)}{k} \left[\cos (\delta_\ell(k,r)) \hat{j}_{\ell}(kr)-\sin (\delta_\ell(k,r)) \hat{\eta}_{\ell}(kr)\right] \\
    &\times\left[\sin (\delta_\ell(k,r))( \hat{j}_{\ell}(kr)+\cos (\delta_\ell(k,r)) \hat{\eta}_{\ell}(kr)\right]
\end{aligned}
\end{equation}
for $\ell=0$ amplitude function becomes
\begin{align}
A_{0}^{\prime}(r) &= -\frac{A_{0} V(r)}{k\left(\frac{\hbar^{2}}{2\mu}\right)} \left[\cos \delta_{0} \cdot \sin(kr) - \sin \delta_{0} \cdot (-\cos(kr))\right] \notag \\
&\quad \times \left[\sin \delta_{0} \cdot \sin(kr) + \cos \delta_{0} \cdot (-\cos(kr))\right]
\label{A0}
\end{align}
also, the equation to obtained wave function is 
\begin{equation}
    u_{\ell}(r)=A_{\ell}(r)\left[\cos (\delta_\ell(k,r)) \hat{j}_{\ell}(k r)-\sin (\delta_\ell(k,r)) \hat{\eta}_{\ell}(k r)\right]
\end{equation}
For $\ell=0$ the wave function takes the following form
\begin{equation}
    u_{0}(r) = A_{0}(r) \left[\cos \delta_{0}(r) \cdot \sin(kr) - \sin \delta_{0}(r) \cdot \cos(kr)\right] 
    \label{u0}
\end{equation}

Higher partial wave amplitude and wavefunction equations ($A_1, A_2, A_3$ and $u_1, u_2, u_3$) can be easily obtained by calculating the the Riccati-Bessel and Riccati-Neumann functions ($\hat{j}_{p},\hat{j}_{d},\hat{j}_{f}$ and $\hat{\eta}_{p},\hat{\eta}_{d},\hat{\eta}_{f}$) respectively. Detailed flowchart to obtain phase shift, amplitude and wave function has been shown in Fig. \ref{flowi}.

\usetikzlibrary{shapes.geometric, arrows}

\tikzstyle{startstop} = [rectangle, rounded corners, minimum width=3cm, minimum height=1cm, text centered, draw=black, fill=red!30]
\tikzstyle{process} = [rectangle, minimum width=3cm, minimum height=1cm, text centered, draw=black, fill=white!30]
\tikzstyle{arrow} = [thick,->,>=stealth]
\tikzstyle{externalarrow} = [thick,->,dashed,>=stealth]

\begin{figure}[h!]
    \centering
    \begin{tikzpicture}[node distance=1.8cm, scale=0.75, transform shape]

    \node (step1) [process] {Interaction Potential;~  $V(r)=-V_0 \text{exp}\{-\alpha r^2\}+1.439975 \frac{Z_1Z_2}{r^2}$};
    \node (step2) [process, below of=step1] {Variable Phase Equation;~  $\delta_{\ell}'(k,r) = -\frac{V(r)}{k\left(\frac{\hbar^{2}}{2\mu}\right)} \bigg[\cos(\delta_\ell(k,r)) \hat{j}_{\ell}(kr) - \sin(\delta_\ell(k,r)) \hat{\eta}_{\ell}(kr)\bigg]^2; \delta_{\ell}(0)=0$};
    \node (step3) [process, below of=step2] {Phase Shift;~  $\delta$};
    \node (step4) [process, below of=step3] {Amplitude Equation;~  $A_{0}' = -\frac{A_{0} V(r)}{k\left(\frac{\hbar^{2}}{2\mu}\right)} \left[\cos \delta_{0} \cdot \sin(kr)-\sin \delta_{0} \cdot (-\cos(kr))\right] \notag \quad \times \left[\sin \delta_{0} \cdot \sin(kr) + \cos \delta_{0} \cdot (-\cos(kr))\right]; A_{\ell}(0)=1$};
    \node (step5) [process, below of=step4] {Amplitude Function;~  $A(r)$};
    \node (step6) [process, below of=step5] {Wavefunction Equation;~  $u_{0}(r) = A_{0}(r) \left[\cos \delta_{0}(r) \cdot \sin(kr) - \sin \delta_{0}(r) \cdot \cos(kr)\right]$};
    \node (step7) [process, below of=step6] {Wavefunction; $u_0(r)$};

    \draw [arrow] (step1) -- (step2);
    \draw [arrow] (step2) -- (step3);
    \draw [arrow] (step3) -- (step4);
    \draw [arrow] (step4) -- (step5);
    \draw [arrow] (step5) -- (step6);
    \draw [arrow] (step6) -- (step7);

    \draw [externalarrow] (step3.east) -- ++(11,0) node[right] {} |- (step6.east);

    \end{tikzpicture}
    \caption{Detailed flowchart for obtaining phase shift, amplitude function, and wavefunction for $p+^7Be$ scattering.}
    \label{flowi}
\end{figure}
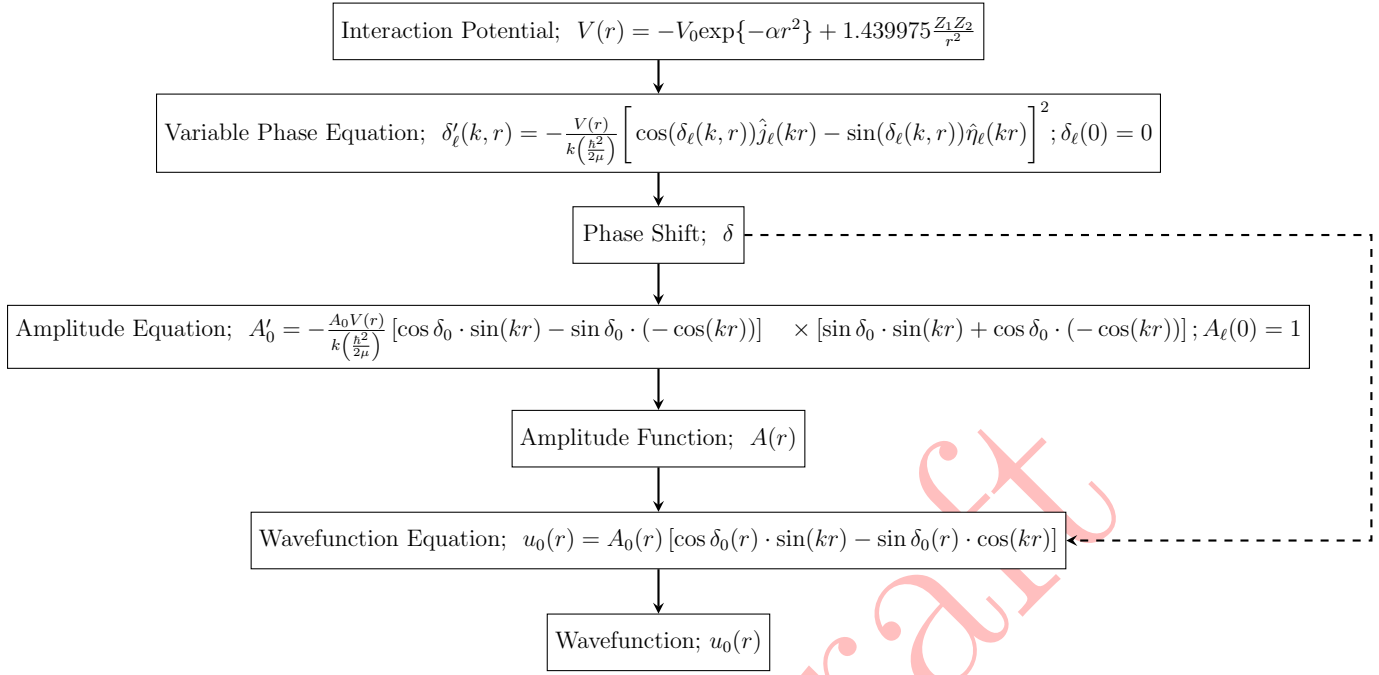

\subsection{Astrophysical S-factor of the $^7Be(p,\gamma)^8B$ capture reaction}
To calculate the astrophysical S-factor one can use the standard expression \cite{1}
\begin{equation}
S(NJ,J_f)=\sigma(NJ, J_f) E_{c.m} \text{exp}\bigg(\frac{31.335 Z_1Z_2\sqrt{\mu}}{\sqrt{E_{c.m}}}\bigg)
\end{equation}
In above equation $\sigma(NJ, J_f)$ is the total capture cross section of the radiative capture via the $EJ$ and $M1$ transitions.

\section{Results and Discussion}
Since the data on the phase shifts for $p^7Be$ elastic scattering  at astrophysical energies is not available, so Dubovichenko et al. Constructed the scattering potential for $p^7Be$ by analogy with $p^7Li$ scattering. Let us consider the structure of the resonance states of $^8B$, which is regarded as a cluster p7Be system:

\begin{enumerate}
\item The first excited level at an excitation energy of 0.7695($\pm$ 2.5) MeV(keV) or \textbf{0.632} MeV in c.m. corresponds to the first resonance state of $^8B$, and a width of 35.6(6) keV in c.m on the spectrum of $^8B$ levels \cite{tilley}. This level can be compared to the sharp resonant $^3P_1$ state of the $p^7Be$ system. We obtained the first resonance energy of 0.565 MeV in c.m. for $^3P_1(R_1)$ state (see table \ref{tab:resonances}) which is closer to the experimentally determined value provided in \cite{tilley}.

\item The second resonance at an excitation energy of 2.32(2) MeV(keV) or \textbf{2.18} MeV in c.m and a width of 350(30) keV (c.m.). It can be matched to the $^3F_3$ state of $p^7Be$ scattering and the $^3F_3$ phase shift has the resonance of 2.523 MeV from our calculations. We consider this transition in order to estimate the value of its cross section.

\item The third resonance level with excitation energy of 3.5($\pm$ 500) MeV(keV) or \textbf{3.66} MeV in c.m, that can be matched to the $^3D_2$ wave, has a width of 8($\pm$ 4) MeV (c.m.). The obtained excitation energy for third resonating state is found to be 3.742 MeV in c.m which is close enough to the experimental value provided by Tilley \textit{et al.} \cite{tilley}.

\item There is second resonance for the $^3P_1(R_2)$ state with an excitation energy \textbf{3.16} MeV at a width of \textbf{1.35} MeV in c.m. The calculated value of second resonance is found to be 3.376 MeV using VPA. 

\item There is a fourth resonance level in the $^8B$ spectrum corresponds to an excitation energy of 10.619(9) MeV(keV) \cite{tilley}, and has not been considered because we are only going upto energy range of 5 MeV in our calculations.

\item For the potential of the bound $^3P_2$ state of the $p^7Be$ system, which refers to the ground state of $^8B$ in the cluster channel under consideration the excitation energy is found to be 3.376 which is quite close to that found in reference \cite{1}. For this potential, a binding energy of -0.137500 MeV \cite{1} is observed i.e., $^8B$ is very weakly bound. The potential depth for $^3P_2$ state is prominently deeper than other states indicating its a bound state relative to other states.
\end{enumerate}

\begin{figure*}
{\includegraphics[scale=1,angle=0]{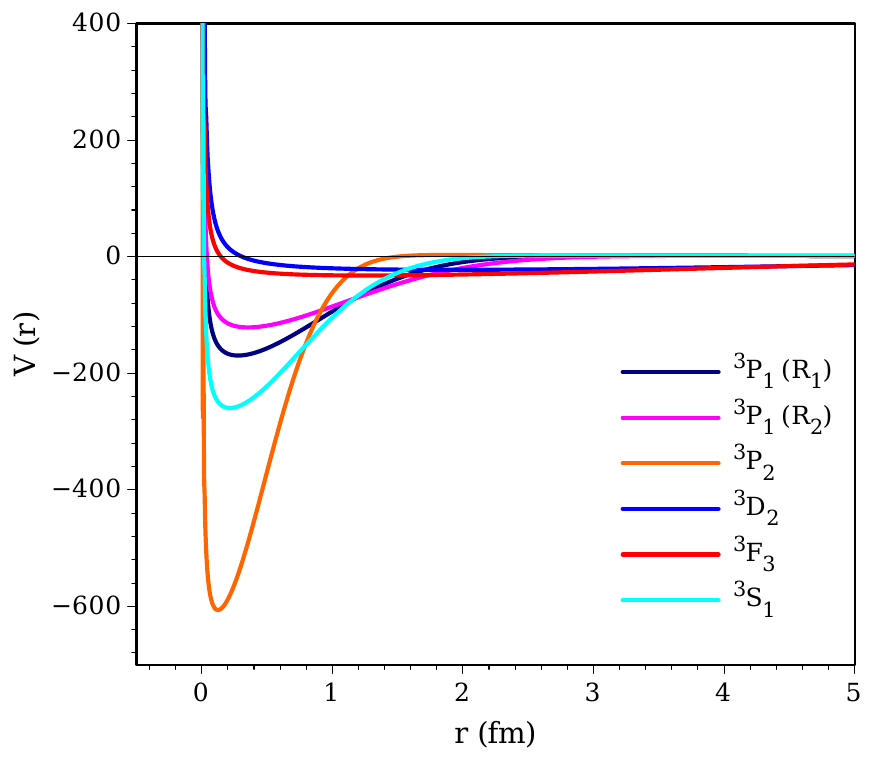}}
\caption{Obtained potentials for different states. The bound state is $^3P_2$ and rest all are scattering states.}
\label{potential}
\end{figure*}

\begin{figure*}
{\includegraphics[scale=0.58,angle=0]{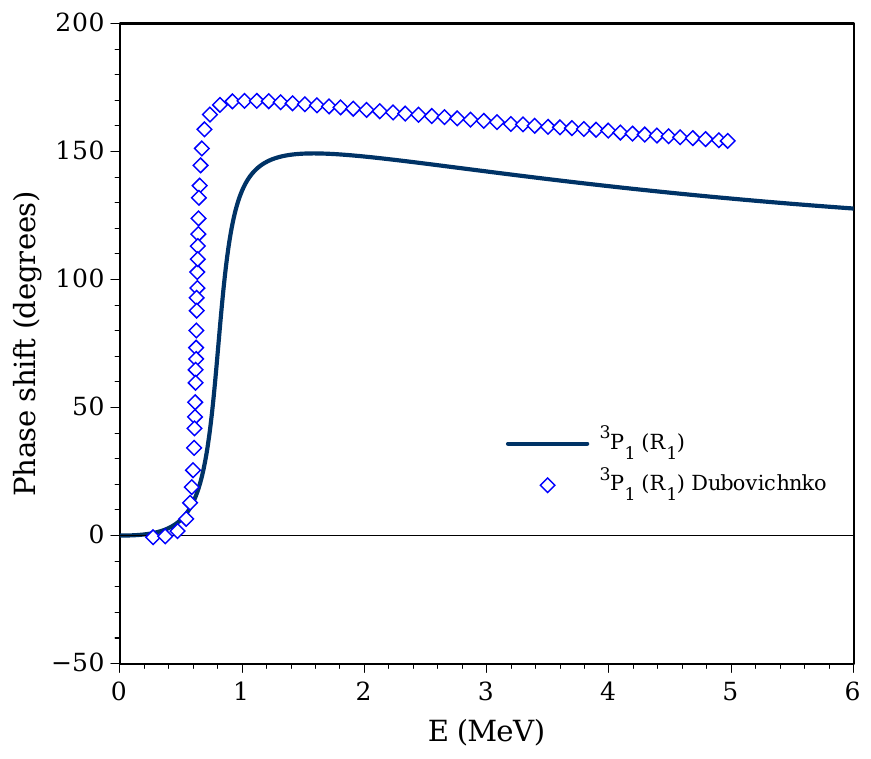}}
{\includegraphics[scale=0.58,angle=0]{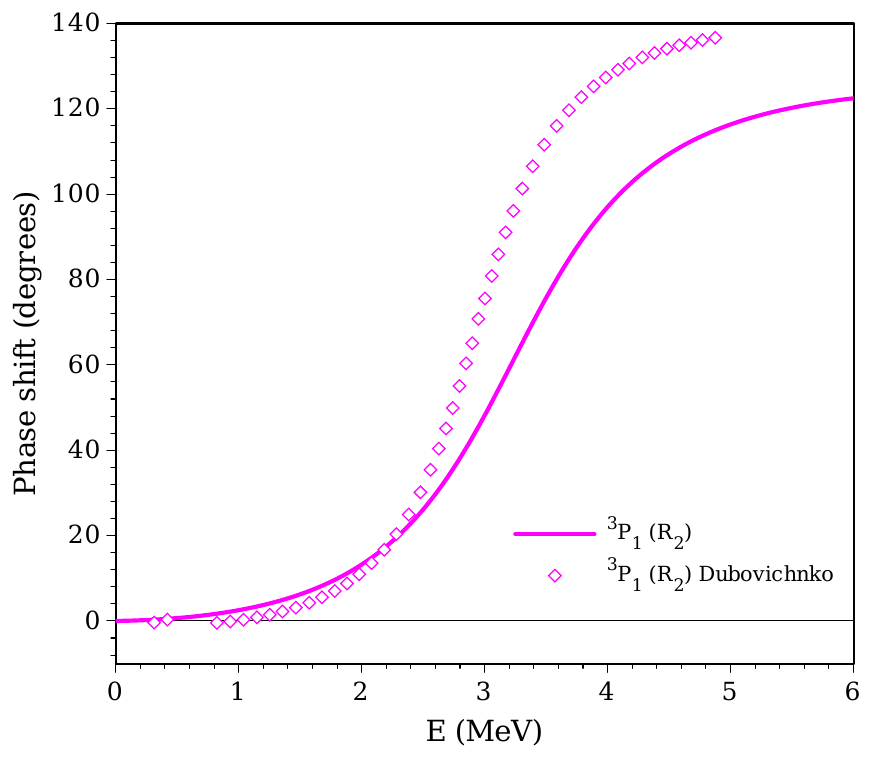}}
{\includegraphics[scale=0.58,angle=0]{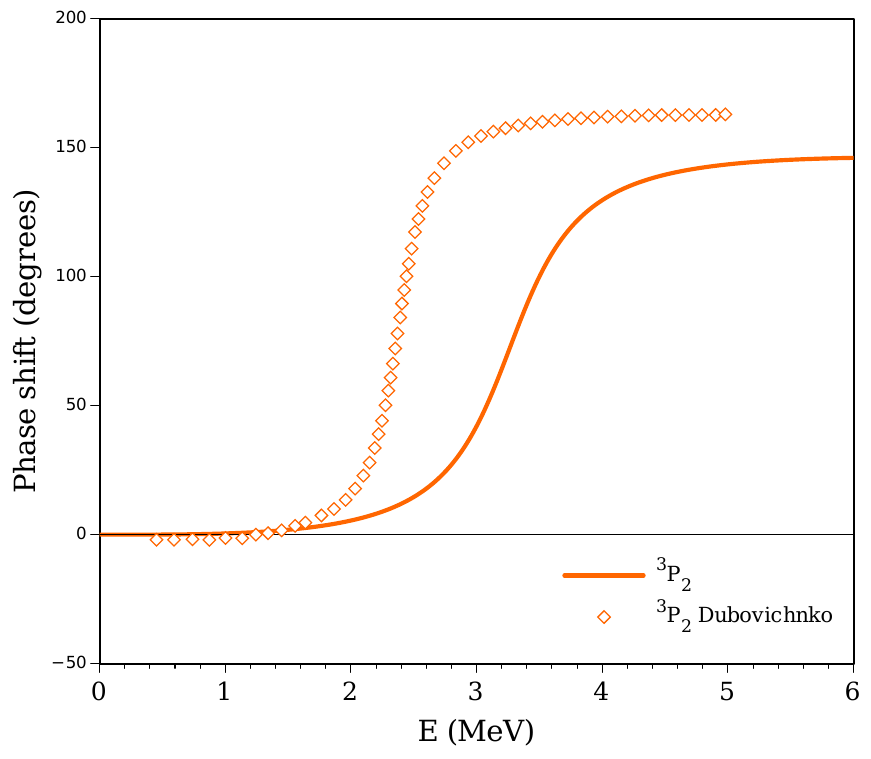}}
{\includegraphics[scale=0.58,angle=0]{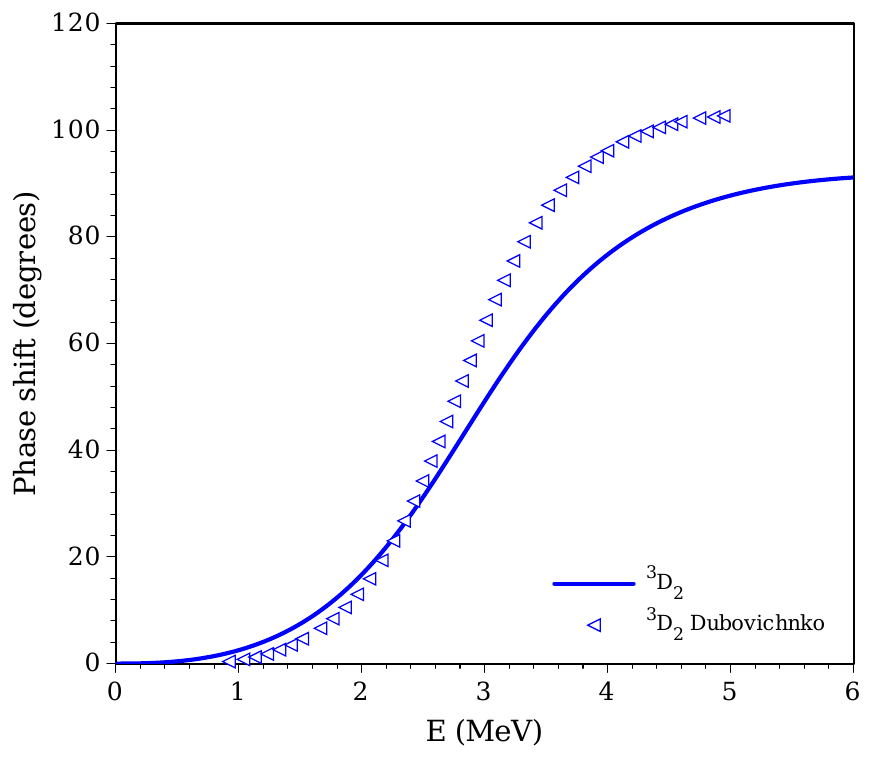}}
{\includegraphics[scale=0.58,angle=0]{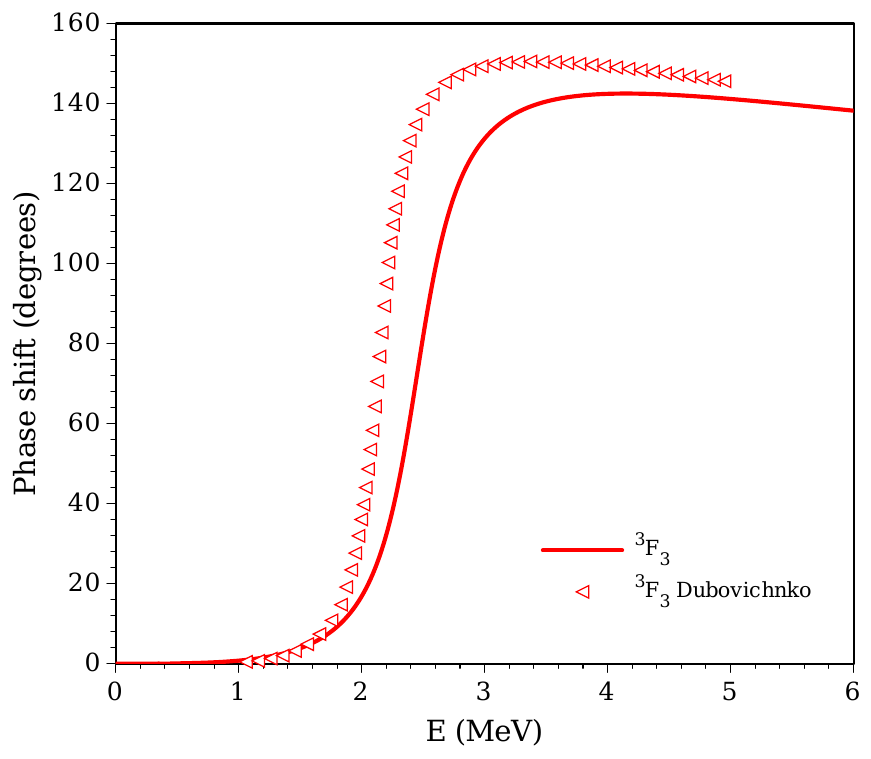}}
{\includegraphics[scale=0.58,angle=0]{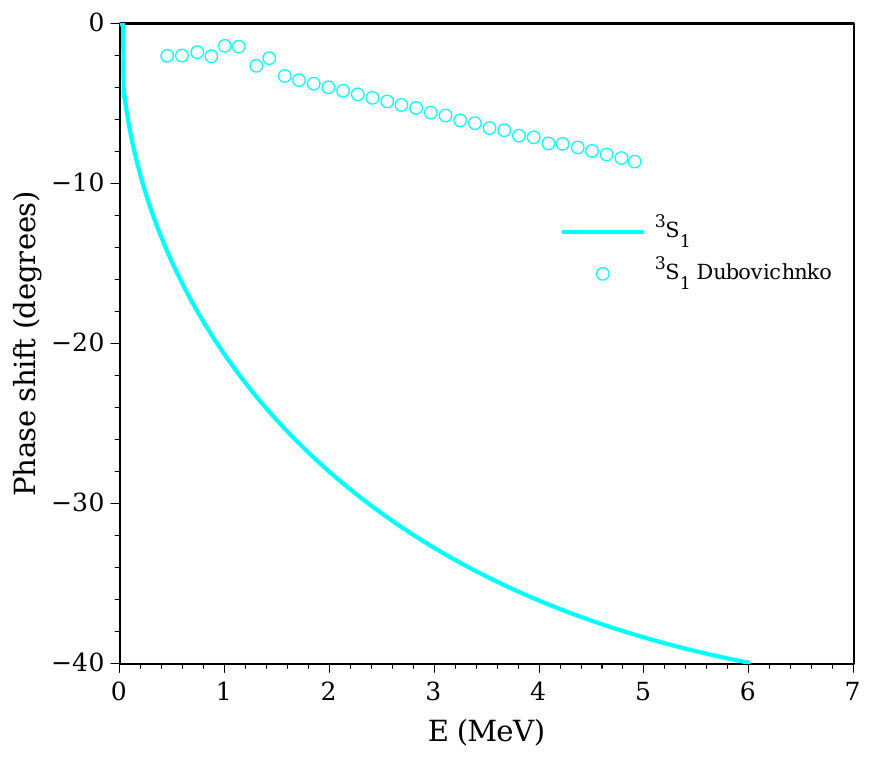}}
\caption{Obtained phase shifts $\delta(E)$ vs \textit{E} for different channels.}
\label{phaseshift}
\end{figure*}

\begin{figure*}
{\includegraphics[scale=0.58,angle=0]{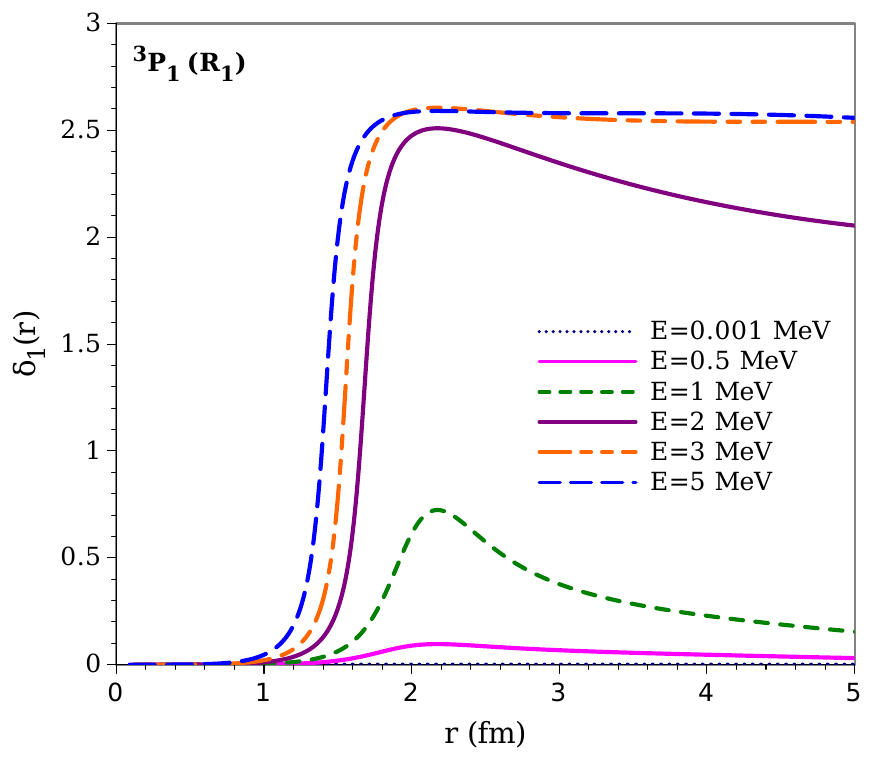}}
{\includegraphics[scale=0.58,angle=0]{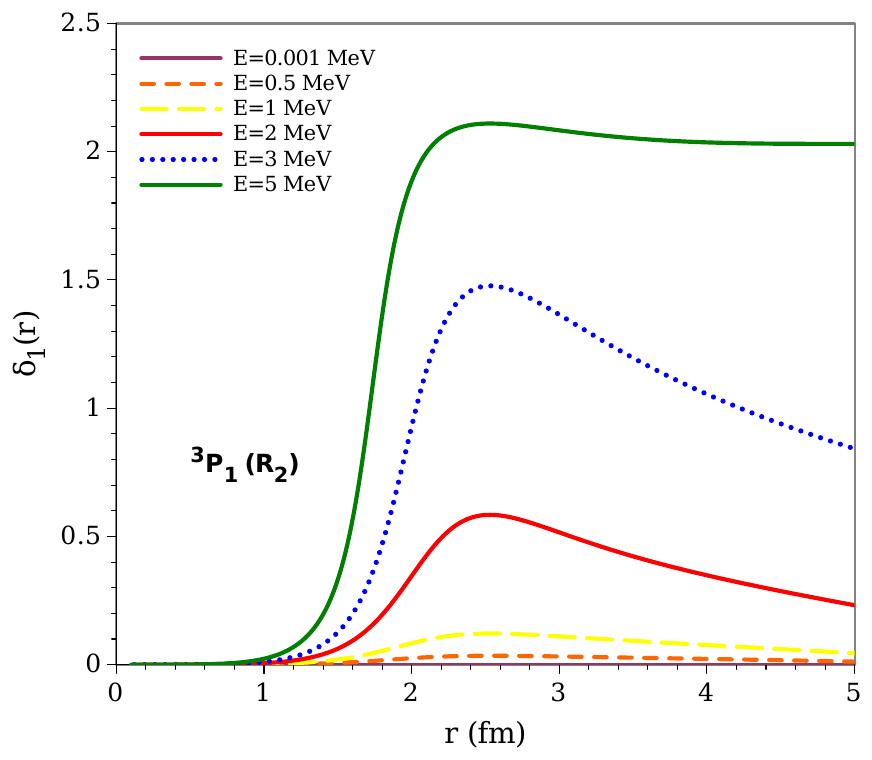}}
{\includegraphics[scale=0.58,angle=0]{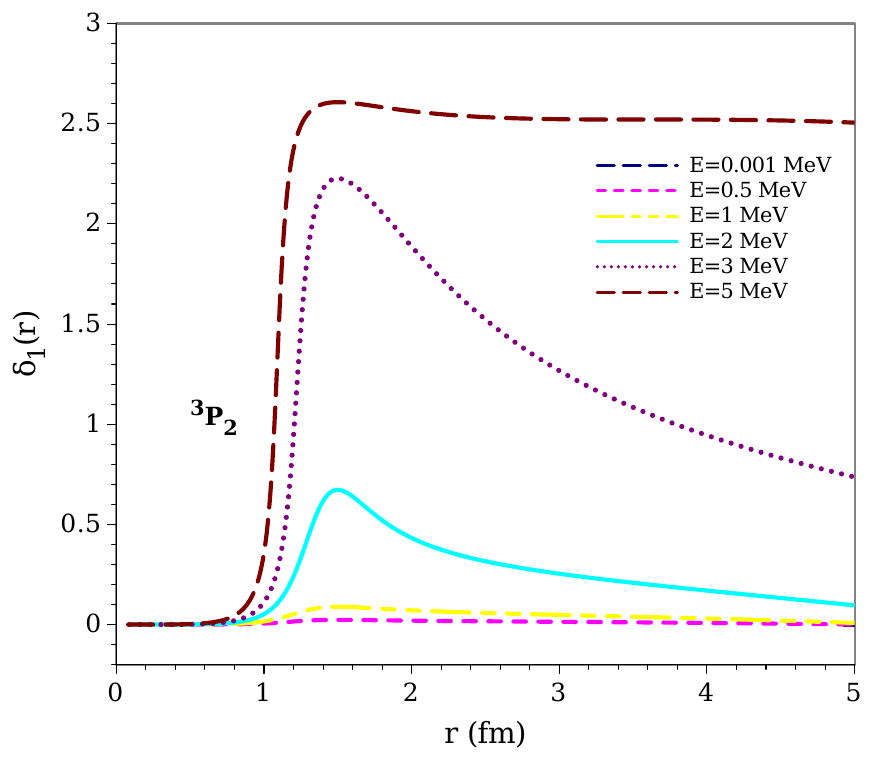}}
{\includegraphics[scale=0.58,angle=0]{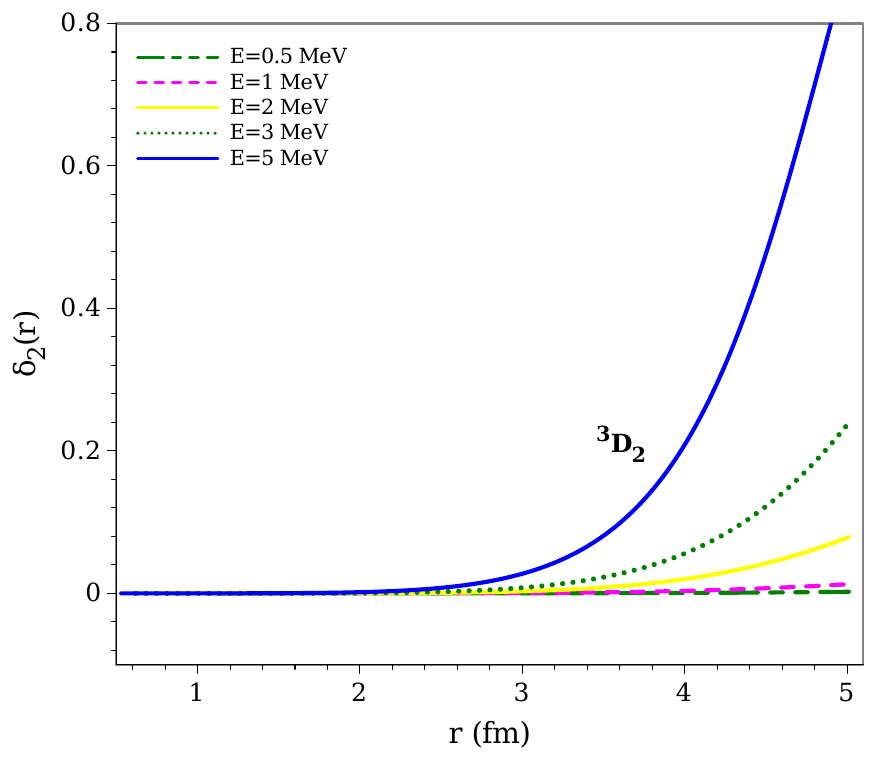}}
{\includegraphics[scale=0.58,angle=0]{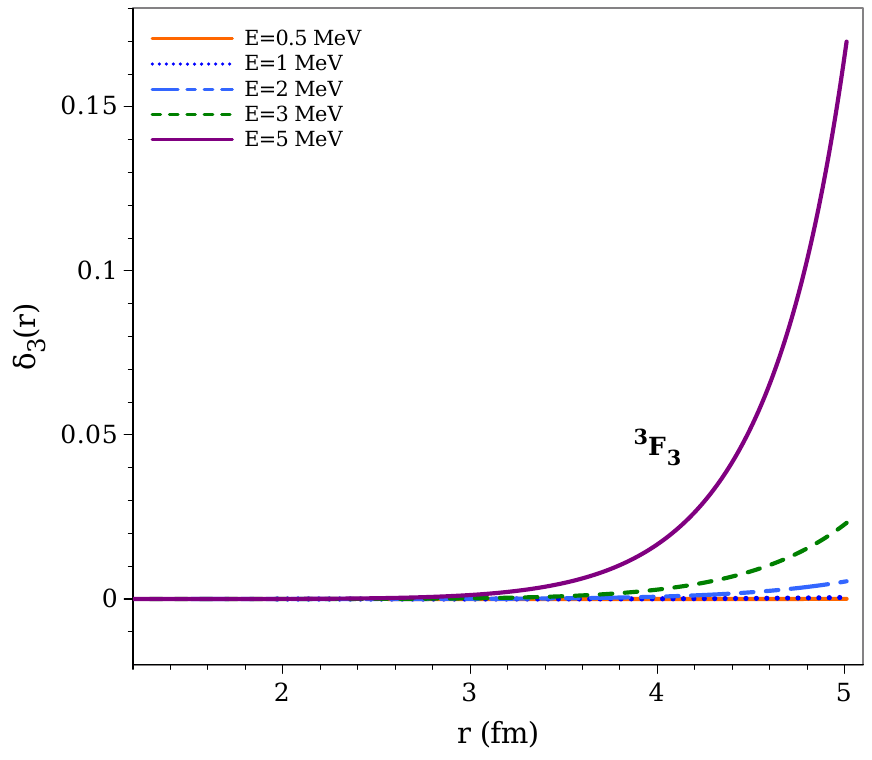}}
{\includegraphics[scale=0.58,angle=0]{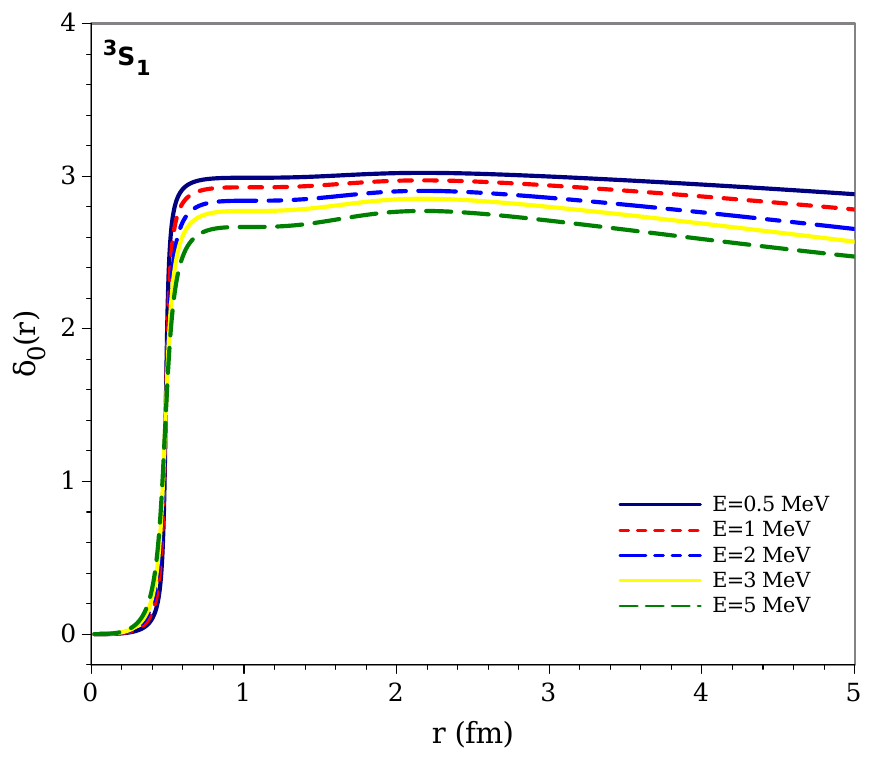}}
\caption{Obtained phase shifts $\delta(r)$ vs \textit{r} for different channels.}
\label{phaseshift_r}
\end{figure*}

\begin{figure*}
{\includegraphics[scale=0.6,angle=0]{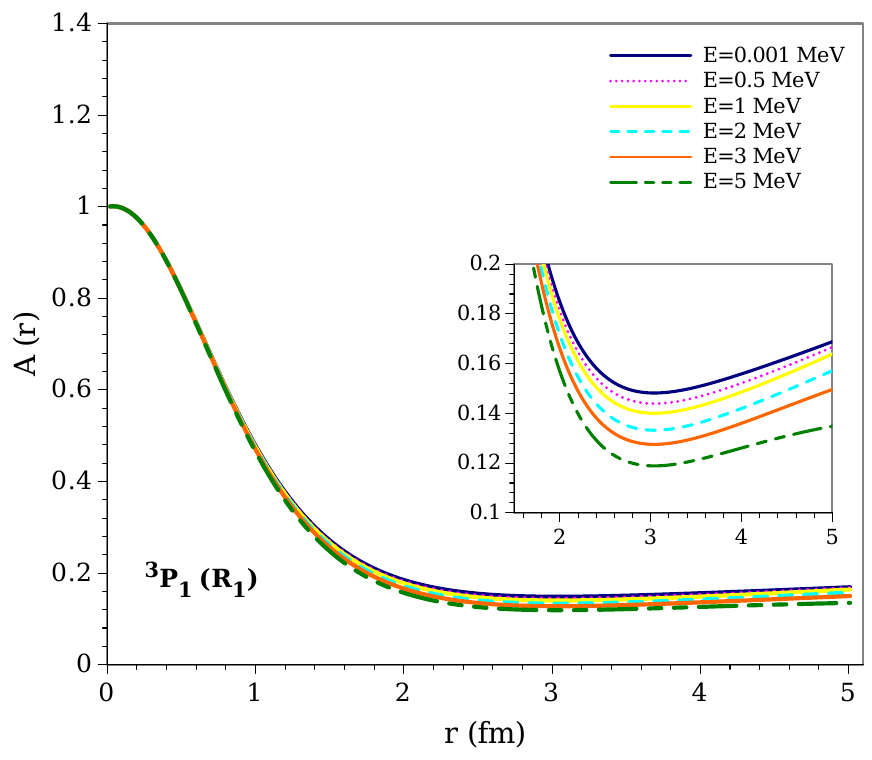}}
{\includegraphics[scale=0.6,angle=0]{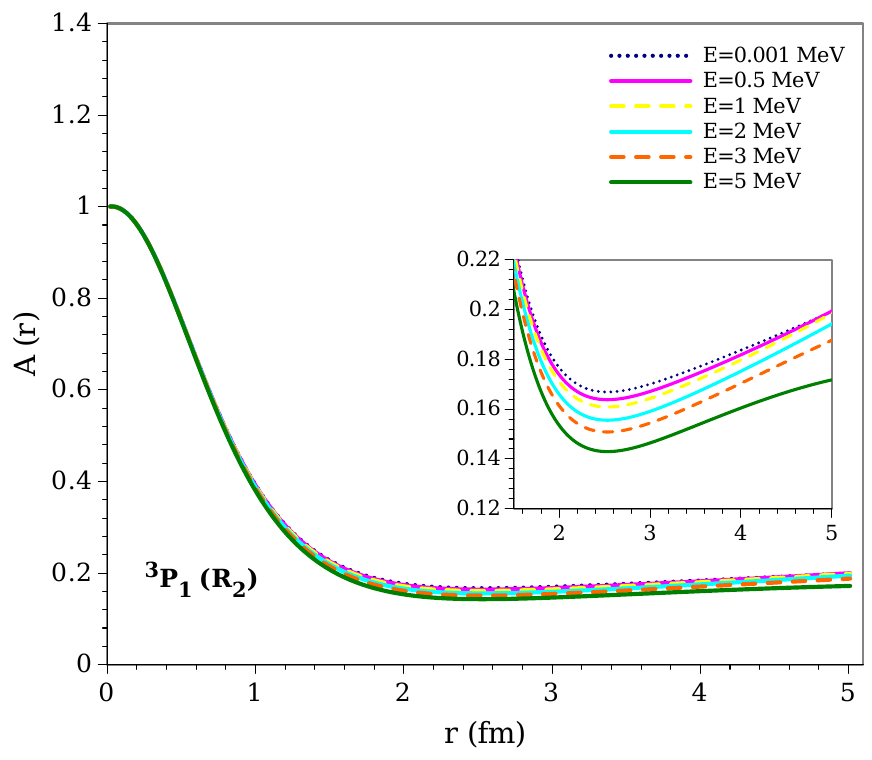}}
\caption{Amplitude function $A(r)$ vs \textit{•}r for $^3P_1 (R_1)$ and $^3P_1 (R_2)$ state.}
\label{amplitude_r}
\end{figure*}

\begin{figure*}
\centering
{\includegraphics[scale=1,angle=0]{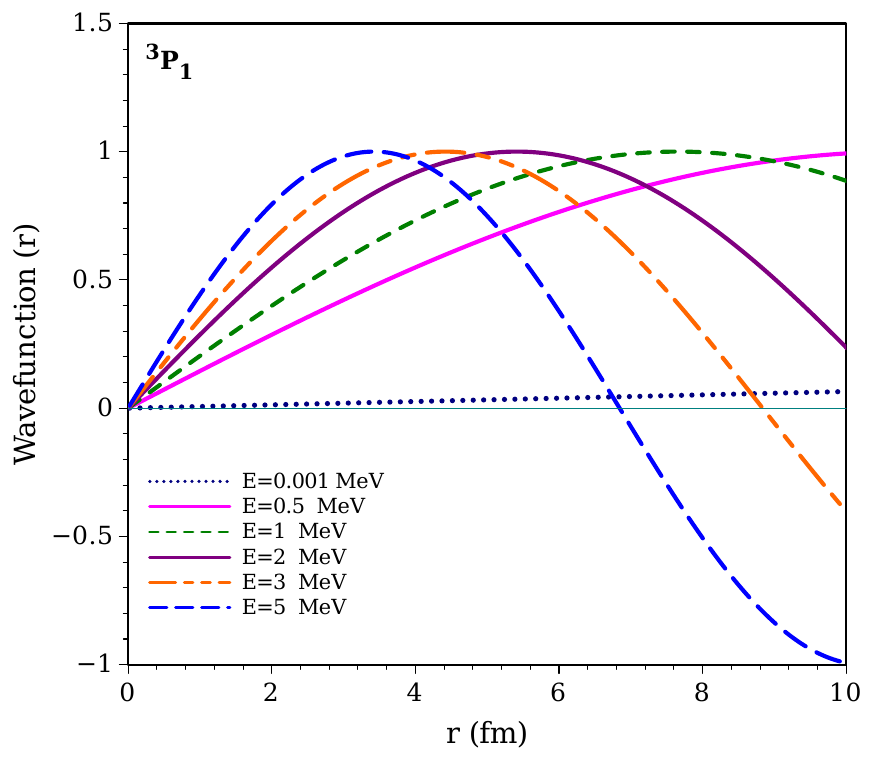}}
\caption{wave function $u_0$ vs \textit{r} for only $^3P_1 (R_1)$ state.}
\label{wf}
\end{figure*}

\begin{figure*}
{\includegraphics[scale=0.58,angle=0]{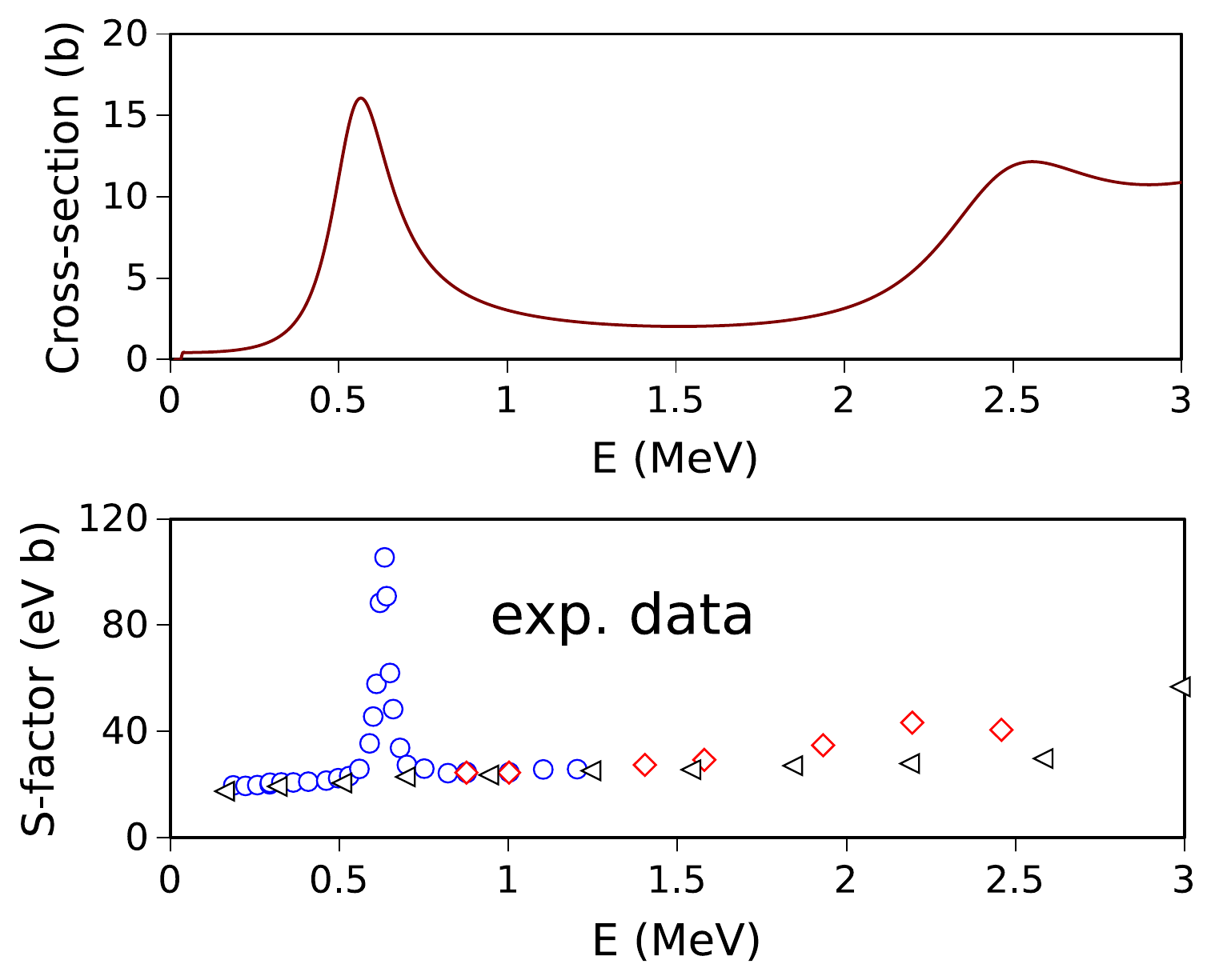} }
\caption{\textbf{(above)} Total partial cross-section plot and \textbf{(below)} experimental S-factor data taken from \cite{2}. In total cross-section peaks are observed at 0.565 MeV while the experimental peak for S-factor is observed at 0.633 MeV.}
\label{tcs_1}
\end{figure*}
In the \( p + {}^7\text{Be} \) system, resonances in the scattering cross section arise due to the formation of a compound nucleus, \( {}^8\text{B} \), in excited states. These resonances correspond to quasi-bound states in \( {}^8\text{B} \), where the incident proton and the \( {}^7\text{Be} \) nucleus form an intermediate system with total energy matching that of an excited level of the compound nucleus. One of the most prominent resonances occurs around \( E_{\text{cm}} \approx 633 \, \text{keV} \), corresponding to an excited state of \( {}^8\text{B} \) at approximately 0.77~MeV excitation energy. This state has been observed to enhance the scattering cross section and contributes significantly to the astrophysical S-factor for the \( {}^7\text{Be}(p,\gamma)^8\text{B} \) reaction, which is a crucial process in solar neutrino production. The presence of such resonances is typically confirmed through phase shift analysis, where a rapid change in phase with energy indicates resonant behavior, particularly in partial waves such as the \( \ell = 1 \) (p-wave) channel. These resonances are characterized by sharp peaks in the cross section. Their inclusion is essential for accurate extrapolation of reaction rates at solar energies, emphasizing the astrophysical importance of understanding resonant structures in low-energy nuclear scattering.

A comparison between the partial cross section and the experimental astrophysical S-factor, as shown in Fig.~\ref{tcs_1}, reveals a clear correlation between the resonance features in the cross section and the enhancements in the S-factor. Notably, both plots exhibit a pronounced structure around \( E \approx 0.6 \, \mathrm{MeV} \), where the partial cross section shows a sharp resonance peak. This resonance coincides with a significant rise in the S-factor, indicating an increased probability of the capture reaction due to the presence of a quasi-bound nuclear state. Additionally, a broader structure is observed near \( E \approx 2.5 \, \mathrm{MeV} \) in both plots, suggesting the influence of higher-energy resonances on the reaction dynamics. These observations underscore the fact that resonant states in the scattering amplitude play a crucial role in determining the energy dependence of the S-factor, particularly in the low-energy regime relevant for nuclear astrophysics.

\begin{figure*}
\centering
{\includegraphics[scale=0.7,angle=0]{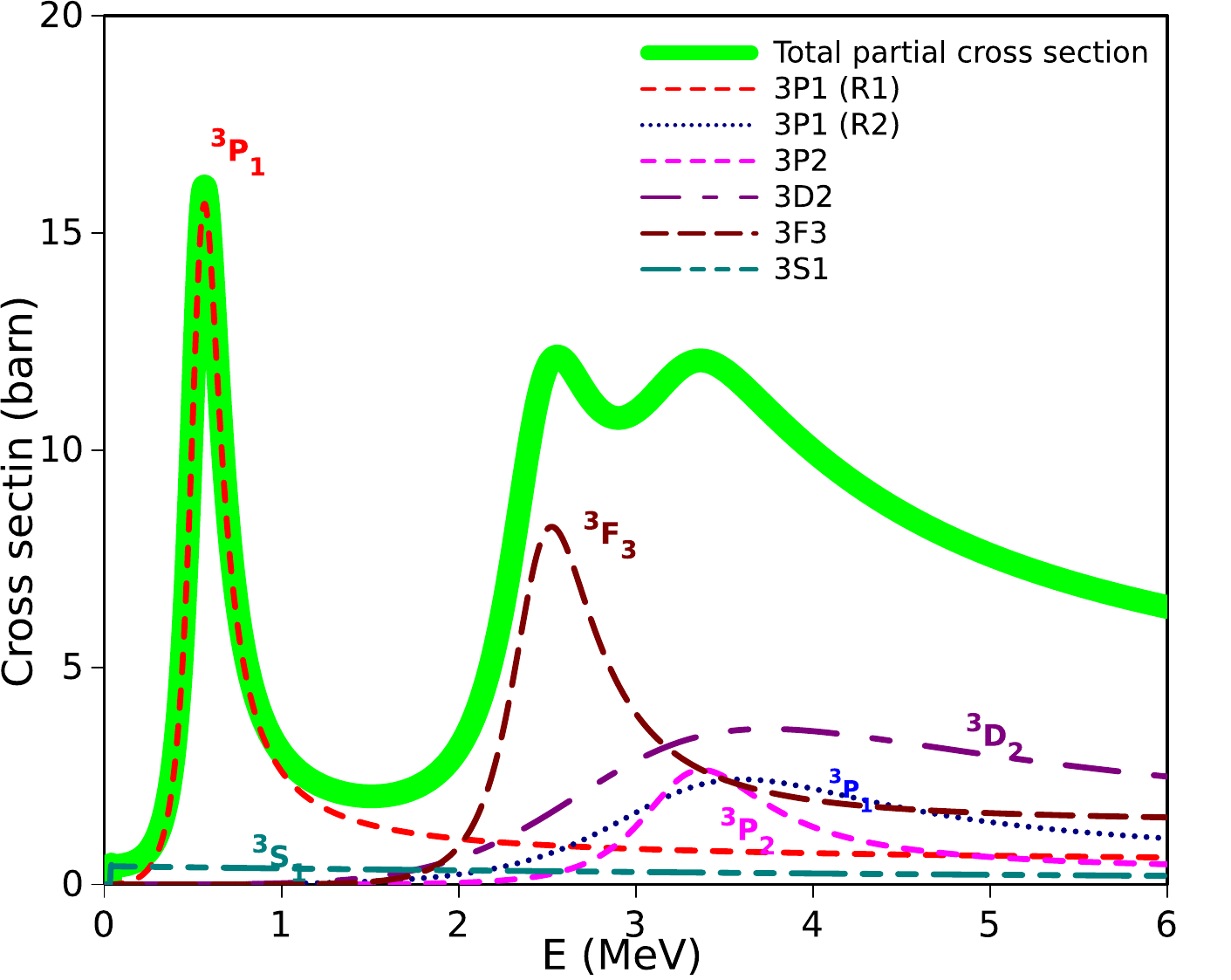}}
\caption{Total and partial cross-section plot for different states. Strong resonance is observed only for $^3P_1$ state while other states shows weak resonance.}
\label{tcs_2}
\end{figure*}

\clearpage
\section{Conclusion}
We have employed the Phase Function Method (VPA) to obtain scattering phase shifts for various channels involved in the $p+^7Be$ system. Using these phase shifts, we further derived important quantum mechanical quantities such as the amplitude function $A(r)$ and the radial wavefunction $u(r)$ for different partial waves. From these, we computed the total partial cross section as the sum over all contributing partial waves. In the present work, we focus on the analysis of the partial cross section for the \( p + {}^7\mathrm{Be} \) system using phase shift information. However, a direct calculation of the astrophysical S-factor has not been performed due to the unavailability of the capture cross section, which is a crucial input for computing the S-factor through the relation \( S(E) = E \, \sigma(E) \, \exp(2\pi \eta) \), where \( \eta \) is the Sommerfeld parameter. The capture cross section involves radiative transition probabilities from the scattering continuum to the bound state of the final nucleus, which requires detailed knowledge of the electromagnetic transition matrix elements and accurate bound-state wavefunctions. As these quantities are not readily available within the scope of the current scattering model, we restrict ourselves to a comparative analysis between the partial cross section and the experimental S-factor data. This comparison nonetheless provides valuable insight, as resonant structures observed in the partial cross section are found to correspond closely with peaks or enhancements in the experimental S-factor, suggesting that the scattering resonances play a significant role in shaping the energy dependence of the capture reaction.

\end{document}